\documentclass[10pt,superscriptaddress, twocolumn,showkeys,aps,prb]{revtex4-1}
\usepackage{cancel}
\usepackage{mathrsfs}
\usepackage{bbold}
\usepackage{float}
\usepackage{graphicx}
\usepackage{bm}
\usepackage{amsmath} 
\usepackage{booktabs}

\def\bsigma{{\pmb{\sigma}}}
\def\btau{{\pmb{\tau}}}

\def\mats{{\mathsf s}}
\def\matbfs{\pmb{\mathsf s}}
\def\nbOne{\pmb{\mathbb 1}}
\def\DeltaSO{\Delta_{\rm SO}}
\def\lambdaSO{\lambda_{\rm R}}

\def\energy{E}

\def\bsigma{{{\sigma}}}
\def\btau{{{\tau}}}

\def\mats{{\mathsf s}}
\def\nbOne{{\mathbb 1}}
\def\DeltaSO{\Delta_{\rm SO}}
\def\lambdaSO{\lambda_{\rm R}}

\def\be{\begin{equation}}
\def\ee{\end{equation}}
\def\beq{\begin{eqnarray}}
\def\eeq{\end{eqnarray}}
\def\half{{\textstyle \frac 12}}

\RequirePackage{color}
\definecolor{MyDarkGreen}{rgb}{0.02,0.60,0.06}

\usepackage{ulem}

\begin{document}

\date{\today}

\title{Persistent charge and spin currents in the long wavelength regime for graphene rings}
\author{N. Bolivar}
\affiliation{Escuela de F\'isica, Facultad de Ciencias, Universidad Central de Venezuela, 1040 Caracas, Venezuela.}
\affiliation{Groupe de Physique Statistique, Institut Jean Lamour, Universit\'e de Lorraine, 54506 Vandoeuvre-les-Nancy Cedex, France.}
\affiliation{Centro de F\'isica, Instituto Venezolano de Investigaciones Cient\'ificas, 21827, Caracas, 1020 A, Venezuela.}
\author{E. Medina}
\affiliation{Centro de F\'isica, Instituto Venezolano de Investigaciones Cient\'ificas, 21827, Caracas, 1020 A, Venezuela.}
\affiliation{Groupe de Physique Statistique, Institut Jean Lamour, Universit\'e de Lorraine, 54506 Vandoeuvre-les-Nancy Cedex, France.}
\affiliation{Escuela de F\'isica, Facultad de Ciencias, Universidad Central de Venezuela, 1040 Caracas, Venezuela.}
\author{B. Berche}
\affiliation{Groupe de Physique Statistique, Institut Jean Lamour, Universit\'e de Lorraine, 54506 Vandoeuvre-les-Nancy Cedex, France.}
\affiliation{Centro de F\'isica, Instituto Venezolano de Investigaciones Cient\'ificas, 21827, Caracas, 1020 A, Venezuela.}

\begin{abstract}
We address the problem of persistent charge and spin currents on a Corbino disk built from a graphene sheet. We consistently derive the Hamiltonian including kinetic, intrinsic (ISO) and Rashba spin-orbit interactions in cylindrical coordinates. The Hamiltonian is carefully considered to reflect hermiticity and covariance. We compute the energy spectrum and the corresponding eigenfunctions separately for the intrinsic and Rashba spin-orbit interactions. In order to determine the charge persistent currents we use the spectrum equilibrium  linear response
definition. We also determine the spin and pseudo spin polarizations associated with such equilibrium currents. For the intrinsic case one can also compute the correct currents by applying the bare velocity operator to the ISO wavefunctions or alternatively the ISO group velocity operator to the free wavefunctions. Charge currents for both SO couplings are maximal in the vicinity of half integer flux quanta. Such maximal currents are protected from thermal effects because contributing levels plunge ($\sim$1K) into the Fermi sea at half integer flux values. Such a mechanism, makes them observable at readily accessible temperatures. Spin currents only arise for the Rashba coupling, due to the spin symmetry of the ISO spectrum. For the Rashba coupling, spin currents are cancelled at half integer fluxes but they remain finite in the vicinity, and the same scenario above protects spin currents. 
\end{abstract}


\maketitle

\section{Introduction}
\indent Graphene probably constitutes one of the most promising materials of the century, not
only because of all its remarkable conduction, topological and mechanical properties, but
also due to its theoretical implications as a testing ground for relativistic effects in low dimensional solid
state systems. In particular, the theoretical applications to spintronic devices are very
promising. As in semiconductors, the presence of Spin Orbit (SO) coupling in Graphene gives a
key element for spin manipulation and is responsible for the existence of the quantum spin Hall phase\cite{Kane2}. 

In a tight binding perspective, the Rashba Spin Orbit coupling (RSO) comes from nearest-neighbour interactions and an applied bias that breaks inversion symmetry, while the Intrinsic SO coupling (ISO) follows from  next nearest-neighbour contributions depending on intrinsic electric fields. The intrinsic interaction is small for free suspended films (1-50$\mu$eV) compared to external perturbations, while the RSO, controlled by an applied bias, can be substantially higher (up to 225 meV) by introducing a coupling to a Ni substrate\cite{Dedkov,Zarea}. The latter enhancement can also be controlled by intercalating Au atoms\cite{Marchenko} between the Ni surface and the graphene film, assuring a more decoupled graphene film and comparable SO strength (100 meV). Intrinsic SO coupling, on the other hand, can be manipulated by functionalizing with heavy atoms on the graphene edges, producing a broad range of parameters where Quantum Spin Hall phases are dominant over electron-electron interaction effects\cite{Autes}. Furthermore, although not graphene based, the same physics has been concocted from ordinary semiconductor superstructures (GaAs), where the SO interaction is much stronger\cite{SushkovNeto} than in suspended graphene. It is then clear that various treatments can be used to build a very significant SO coupling into graphene ribbons and rings giving a starting point for spintronics based device concepts. 

Graphene quantum rings have recently attracted much attention for many reasons among which we mention: i) Confining Dirac fermions is non-trivial because of Klein's paradox. Various mechanisms have been devised to overcome reduced backscattering from scalar potentials, such spatially modulating finite Dirac gaps\cite{Zhu} or through spatially inhomogeneous magnetic fields\cite{Egger}. A direct approach is simply mechanically cutting\cite{flakecuttingpaper} into confining geometries creating and infinite mass boundary. ii) The multiply connected structure of the ring gives rise to Aharonov-Bohm oscillations in external fields\cite{Recher} that can be manipulated by effective gauge fields generated through strain\cite{Faria}. iii) Both ferromagnetic (FM) and antiferromagnetic (AF) phases exist, when contemplating electron-electron and/or spin-orbit interactions,  that live on the graphene edges, and their magnitudes are enhanced in ring geometries\cite{Grujic}. iv) Rings in Mobius topologies induce spin Hall effect in graphene and various FM and AF phases, even without SO couplings when electron-electron interactions are considered. v) Persistent currents are a ground state phenomenon induced by time reversal symmetry breaking and manifest themselves as a ground state current in coherent conditions. Ring confinement in graphene has been shown to lead to controlled lifting of the valley degeneracy in conjunction with a magnetic flux\cite{Recher}. The footprint of this broken valley degeneracy is a charge persistent current.

In this work we address graphene rings where Dirac fermions are confined by an infinite mass barrier, by either growing the ring epitaxially\cite{chinesejapaneseepitaxial} or cutting it out by chemical means\cite{flakecuttingpaper}. We consider the bare Dirac Hamiltonian plus either the Rashba or the intrinsic SO couplings. The former can be modulated by either gate voltages or charge transfer to a contrasting substrate causing large perpendicular electric fields. The intrinsic coupling enhanced by e.g. edge heavy atom functionalization as discussed before. Under these conditions we compute the spectrum and the eigenfunctions in the ring geometry, for large enough rings, so that boundary conditions can be considered as zigzag to a high degree of approximation\cite{Beenakker}.  We assume that the ring is in the lowest radial state, and no mixture occurs with higher excited states\cite{Shakouri}, a consideration that we will show is warranted. Finally we ignore electron-electron interactions and consider the SO coupling is dominant\cite{Autes}. 

\section{The graphene ring model}
\begin{figure}
\includegraphics[scale=0.39]{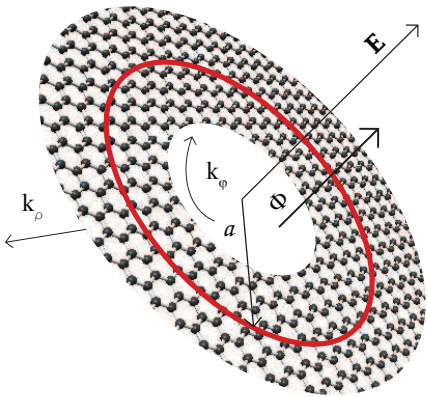}
\caption{A quasi one-dimensional graphene ring in the long wavelength approach inherits all the internal symmetries the lattice structure. The Corbino ring is cut out from a graphene monolayer. The figure shows the ($\rho,\varphi$) coordinates, $z$ being perpendicular to the plane. The average radius of the ring is $a$, represented by the red line. The ring is pierced by a perpendicular magnetic flux and electric field.}
\label{fig:figuraNT}
\end{figure}

A quasi one-dimensional ring (Corbino disk) of finite width, is cut out from a flat graphene monolayer as shown in Fig.\ref{fig:figuraNT}. Localized and discrete confined modes exist in the radial direction (see \cite{Fertig} for the case of carbon nano ribbons), while the angular direction is free, though appropriate closing conditions for the wave functions are to be applied.
The starting Hamiltonian is given by that of a graphene sheet\cite{huertas} with SO interaction in the long 
wavelength limit around the Dirac points, 
\begin{eqnarray}
H&=&-i\hbar v_F (\btau_z \bsigma_x  \nbOne_{s} \partial_x+\nbOne_{\tau} \bsigma_y  \nbOne_{s} \partial_y)\nonumber \\
&& + \DeltaSO \btau_z \bsigma_z  \mats_z + \lambdaSO(\btau_z \bsigma_x  \mats_y - \nbOne_{\tau} \bsigma_y  \mats_x ).
\label{ham1}
\end{eqnarray}
The first term is the kinetic energy, it has the form $v_F {\bm\sigma} \cdot {\bf p}$, with the additional $\btau_z-$Pauli matrix which acts on the ``valley'' index and distinguishes between the Dirac points in the band structure, ${\bf k}=\tau{\bf K}=\tau(4\pi/3c,0)$ where $c$ is the distance between the Bravais lattice points and $\tau$ takes on values of $\pm 1$. The $\bsigma_i-$Pauli matrices encodes for the sublattice distinction. The $\mats_i$ represents the real spin of the charge carriers. Products of matrices in the Hamiltonian are understood as tensor products between different sub-spaces.
When only two operators or less are present, identity $2\times 2$ matrices
is implied for each of the omitted subspaces.

The second term in Eq.\ref{ham1} is the intrinsic spin orbit (ISO) coupling\cite{Kane}, due to the electric fields of the carbon atoms. From a tight binding point of view, this interaction comes from second neighbor hopping contribution that preserves all the symmetries of graphene.  The last term in Hamiltonian Eq.\ref{ham1} is the Rashba SO interaction which results from the action of an external electric field that breaks the space mirror symmetry\cite{MinEtAl} with respect the graphene plane. 
    
 The operators $\sigma_i$ and $\mats_i$ are dimensionless and normalized as $\sigma_i^2=\nbOne_\sigma$ and $\mats_i^2=\nbOne_s$ (Pauli matrices), and the parameters $\DeltaSO$ and $\lambdaSO$ have dimensions of an energy. 
Since the valley operator $\tau_z$ is diagonal, the Hamiltonian can be split into two different contributions, one for each valley in $\bf{k}$ space, thus reducing the model to two copies of a $4 \times 4$ matrix system, instead of the full $8 \times 8$ direct product space. One then gets two separate valley Hamiltonians, 
\begin{eqnarray}
H_{+}&=&-i\hbar v_F (\bsigma_x \partial_x+\bsigma_y \partial_y)\nonumber\\
&&\quad+ \DeltaSO \bsigma_z  \mats_z +   
 \lambdaSO(\bsigma_x \mats_y - \bsigma_y \mats_x ),\label{eqH+}\\
H_{-}&=&-i\hbar v_F (-\bsigma_x \partial_x+\bsigma_y \partial_y) \nonumber\\
&&\quad- \DeltaSO \bsigma_z \mats_z -
 \lambdaSO(\bsigma_x \mats_y + \bsigma_y \mats_x ). \label{eqH-}
\end{eqnarray}
so each valley can be treated separately.

\section{Ring Hamiltonian, Boundary Conditions and Hermiticity}

In this section we will clarify a few important general points for the Hamiltonian in polar coordinates which are frequently overlooked in the literature. The first concerns the boundary conditions imposed on the wave function in a non-simply connected geometry, and the other, the form of the Hamiltonian used when a change in coordinate system is involved. We will first derive the closed ring Hamiltonian of radius $a$ of pure kinetic energy by performing the proper coordinate mapping\cite{Berch}. Then we will discuss the boundary conditions (BCs) on the ring geometry. The salient features of this model relevant to the full Corbino disk Hamiltonian will be transparent. 

When using coordinates other than Cartesian, one must take care of subtleties in constructing an hermitian Hamiltonian\cite{morpurgo}, whose correct form avoids spurious features in the spectrum. In the $\tau=1$ valley, keeping only the kinetic energy and omitting the spin degree of freedom, the coordinate change applied to Eq.~(\ref{eqH+}) results in 
\begin{eqnarray}
H=-i \frac{\hbar v_F}{a}(\bsigma_y \cos\varphi\ \! - \bsigma_x \sin\varphi\ \!)\partial_\varphi,\label{eq-HnonHermitian}
\end{eqnarray}
after removing the radial part. The difficulty comes from the observation that this Hamiltonian is not hermitian\cite{Cotaescu}, since 
$\langle F \mid H \mid G \rangle^* \neq \langle G \mid H \mid F\rangle$,
where $\mid\! F\rangle$ and $\mid\! G\rangle$ are 2-components spinors. This can be repaired by adding to 
Eq.~(\ref{eq-HnonHermitian}) a term
proportional to $i (\bsigma_y \sin\varphi\ \! + \tau \bsigma_x  \cos\varphi)$~\cite{Berch} and one is easily led to
the form,
\begin{eqnarray}
H_{\tau}=&-&i\frac{\hbar v_F}{a} [ ( -\tau \bsigma_x  \sin\varphi\ \! + \bsigma_y \cos\varphi\ \! )\partial_\varphi +\nonumber \\
&-&  \half(\bsigma_y \sin\varphi\ \! + \tau \bsigma_x  \cos\varphi\ \!)]\label{eq_HamHerm}.
\end{eqnarray}
Not including this term would also lead to real, but physically incorrect eigenvalues and 
eigenstates\cite{Cotaescu}. The reason for real eigenvalues, in spite of non-hermiticiy, follows from the
operator being $PT$ (parity and time reversal) symmetric\cite{BenderEtAl}.

Another way of deriving the correct form for the Hamiltonian is the following: Imagine that we start with the Hamiltonian
$H=v_F{\bm \sigma}\cdot{\bf p}$. Writing it directly in polar coordinates, one gets
\beq H&=&-i\hbar v_F\Bigl(\bsigma_\rho \partial_\rho +\rho^{-1}\bsigma_\varphi\partial_\varphi\Bigr),\label{eq3d}\eeq 
fixing $\rho=a$ and taking care to properly symmetrize the product  $\bsigma_\varphi\partial_\varphi$, one describes a ring 
\be H_{\rm ring}=-i\hbar v_Fa^{-1}(\bsigma_\varphi\partial_\varphi-\half\bsigma_\rho),\label{eqRing1d}\ee 
that corresponds to the expression in Eq.\ref{eq_HamHerm} using $\bsigma_\rho=\bsigma_x\cos\varphi+\bsigma_y\sin\varphi$ and 
 $\bsigma_\varphi=-\bsigma_x \sin\varphi+\bsigma_y\cos\varphi$.
The term  $\half\bsigma_\rho$ in Eq.~\ref{eqRing1d}  is essential since it renders
the derivative in polar coordinates covariant by introducing the connection that correctly rotates the internal
degree of freedom so as to keep the pseudo spin parallel to the momentum.  This form of the Hamiltonian
for the angular dependence is arrived at independently of the form of the confining potential applied
radially\cite{morpurgo,Fertig}. The details of the confining potential will arise as effective coefficients in the
form of Eq.\ref{eqRing1d}.

The eigenstates of Eq.\ref{eqRing1d}
are of the form 
\begin{equation}
\psi(\varphi,z)=\frac{ e^{im\varphi}}{\sqrt{2}}\begin{pmatrix} -i\kappa e^{-i\varphi}\\  1\end{pmatrix},
\end{equation}
with $m$ a half positive integer for metallic rings and $\kappa=\pm 1$ describing electrons and holes respectively. The corresponding energies are $E=\frac{\kappa\hbar v_F}{a} (m-1/2)$. It is easy to verify that $\langle {\bm \sigma}\rangle=\kappa(-\sin{\varphi},\cos{\varphi})=\kappa\bf{k}/|\bf{k}|$, as can be derived in Cartesian coordinates. In spite of the fact that pseudo-spin follows the momentum, it is endowed
with proper angular momentum\cite{LessonsGraphene}. This can be verified by noting that  the Hamiltonian of Eq.\ref{eqRing1d} does not commute with the orbital angular momentum alone $L_z=-i\hbar\partial_{\varphi}$ but with the combination $J_z=L_z+\frac{1}{2}\hbar\sigma_z$ (and with $(L_z{\bf u}_z+\frac 12\hbar\bm \sigma)^2$. Note that if $L_z$ does not commute with the Hamiltonian, there is a torque on the orbital momentum. This torque is compensated by a torque on the pseudo spin angular moment so that the total $J_z$ is conserved. Thus, with the pseudo spin there is
associated ``lattice spin" presumably from the rotation the electron sees of the A-B bond. We will see in the next section, how this extra angular momentum 
combines with the regular electron spin to generate a total conserved angular momentum.

The wave function nevertheless, preserves spin-like properties. One can verify that the the ring eigenfunctions are anti-periodic, thus  $\psi(\varphi+2\pi)=-\psi(\varphi)$, a property which finds its origin in the effect of the $2\pi$ rotation on the connection $\frac{1}{2}\sigma_\rho$. The factor $\half$ corresponds to a Berry phase, discussed as a very crucial feature of graphene (e.g. by Katsnelson~\cite{Katsnelson}, Guinea et al~\cite{CNeto}) and of carbon nanotubes (e.g.  in Ref.~\onlinecite{Ando98}). We will reemphasize these points in our derivation in the following sections which several previous references have overlooked (see references \onlinecite{Cotaescu,GonzalezEtAl,Recher,Berch}).

\section{Closing the wave function on a graphene ring}
In this section, we are interested in discussing graphene rings described with the effective Dirac theory in the vicinity of the $K$ points with appropriate boundary conditions. Recalling that according to Bloch's theorem, the wave function $\psi({\bf r})=u_{{\bf k}}(\bf r)e^{i{\bf k}\cdot{\bf r}}$  should exhibit the ring periodicity, while the Bloch amplitude $u_{\bf k}(\bf r)$ has the lattice periodicity. The periodic boundary conditions imposed on the Bloch wave function do not necessarily imply periodic boundary conditions for the eigenfunctions
of the effective theory~\cite{KaneMele97}. Indeed, $\bf k$ is measured from the Brillouin zone 
center ($\Gamma$ point). The effective theory is related to the wave vector ${\bf q}={\bf p}/\hbar$ 
in the neighborhood of the Dirac points through ${\bf k}={\bf K}_D+{\bf q}$. 

Generalizing the boundary conditions for the case of a ring with linear dispersion (see previous section) we introduce a twist phase $\theta_0$ in the closing of the wave function 
\be
\psi(\varphi+2\pi)=e^{-i\theta_0}\psi(\varphi)\label{eq-twistedBC}.
\ee
The eigenstates are now of the form 
\be
\psi(\varphi)=e^{i(m-\theta_0/2\pi)\varphi}\begin{pmatrix}Ae^{-i\varphi}\\ B\end{pmatrix},
\ee
with $m$ an integer, with corresponding eigenvalues
\be
\energy=\pm\frac{\hbar v_F}{a}\Bigl|
m-{\textstyle\frac {\theta_0}{2\pi}}
\Bigr|
,\label{EqEnergiesTwistedGraphene}
\ee
where $\kappa=\pm 1$ refers to particles (conduction band) and holes (valence band).
As we discussed in the previous section, in the case of a graphene ring, antiperiodic BCs (ABC) should be chosen\cite{KaneMele97}; this means $\theta_0=\pi$ for graphene in a Corbino geometry, but for different boundary conditions, such as those that occur in carbon nanotubes with arbitrary chiralities, can also be described. Note that the twist phase plays the same role as a magnetic flux through the ring, that can modify its conducting properties by manipulating the gap at the Dirac point. 

It is important to discuss the boundary conditions on the graphene rings that we will consider.
As a reference, graphene nano ribbons have been addressed in detail\cite{Enoki}. For the approximation addressed here, the zig-zag nano-ribbons are the closest relative, since it has been shown\cite{Beenakker} that a generically cut honeycomb lattice has approximately zig-zag boundary conditions to a high accuracy.  Once zig-zag boundaries are assumed it has been shown\cite{Fertig} that a continuum Dirac description can well approximate nanoribbons modelled by the tight-binding approximation with less than $1\%$ error at least for widths of 10 times the basis vector length. The continuum description gets better as the nanoribbon increases width.

For graphene ribbons with zigzag edges there is the concern that longitudinal and transverse states are coupled\cite{Enoki} and slicing the graphene band using the boundary conditions is not warranted for small ribbon widths $N\sim 1$ (number of transverse lattice sites). Nevertheless, for wide ribbons ($N\gg1$) this approximation becomes increasingly good as  can be judged from the relation coupling the longitudinal $\rm k$ and transverse $\rm p$ modes $\sin{{\rm p}N}+w\cos({\rm k}/2) \sin{{\rm p}(N+1)}=0$ where the wavectors in units of the magnitude of the primitive translation vectors of the lattice. When $N\gg1$ then ${\rm p}=m\pi/N$ independent of $\rm k$. One final concern is the existence of one localized state that for nano ribbons for a critical value of the longitudinal wavevector, nevertheless, the restriction also disappears in the limit $N\gg1$ in which our continuum approximation is based.  In the next section we will discuss the possible coupling of the transverse modes due to the spin-orbit interaction.

The vicinity to the Dirac points is an important issue here, since the linear range of the spectrum is subject to the lattice parameter and the radius of the ring. The estimated limiting value of the momentum ignoring lattice effects\cite{Sarma} is $k_l \approx 0.25 {\rm nm^{-1}}$. The carrier limiting energy at this point is $E_l=\hbar v_F k_l$. Equating this value with Eq. (\ref{EqEnergiesTwistedGraphene}) we obtain the maximum number of states, hence, 
\be
\Bigl| m-{\textstyle\frac {\theta_0}{2\pi}}
\Bigr| \lesssim k_l a. 
\ee
As a reference estimation based on an analogous ring already present in nature (in fact a carbon nanotube section has a kinetic term of the same form as the Corbino); a single wall carbon nanotube has radius that goes from $ 10~ {\rm nm}$ to $100~ {\rm nm}$, this gives order of magnitudes from $m \sim 2$ to $m \sim 25$ that varies depending whether it is an armchair or zigzag tube. For a carbon nanotube with a smaller radius than $4~{\rm nm}$, the allowed states will be outside the linear region establishing a threshold for the values of $a$ in the long wavelength approach.

\section{Spin-Orbit coupling}
Having set up the correct Hamiltonian and boundary conditions to describe a graphene ring, we can incorporate SO interactions in the ring geometry to obtain the equivalent of Eqs.~(\ref{eqH+}) and (\ref{eqH-}). The spectrum becomes independent of the valley index $\tau$, so we will only deal with $\tau=+1$ in polar coordinates:
\beq
 H_{+}&=&-i\hbar v_Fa^{-1}(\bsigma_\varphi\partial_\varphi-\half\bsigma_\rho)\nonumber\\
 &&\quad+ \DeltaSO \bsigma_z  \mats_z +   
 \lambdaSO(\bsigma_\rho \mats_\varphi - \bsigma_\varphi \mats_\rho ).
  \label{eqRing1d+}
\eeq
For the ISO only case we assume a $4-$component vector to represent the electronic states, incorporating electron spin, 
$\Psi=e^{im\varphi}\left (A^{\kappa,\delta}_\uparrow e^{-i\varphi},A^{\kappa,\delta}_\downarrow, B^{\kappa,\delta}_\uparrow, B^{\kappa,\delta}_\downarrow e^{i\varphi}\right )^T$, where $A^{\kappa,\delta}_{\uparrow,\downarrow}~(B^{\kappa,\delta}_{\uparrow,\downarrow})$ is the wavevector amplitude on sublattice $A$ ($B$) with spin $\delta={\uparrow\downarrow}$. 
The ansatz for the spinor is constructed in order to account for the conservation of the total angular momentum. All components carry the same angular momentum $J_z$, adding in units of $\hbar$ a purely orbital contribution (respectively $m-1$, $m$, $m$, and $m+1$ for the four compoents), a
pseudo-spin or lattice contribution (resp. $+\frac 12$, $+\frac 12$, $-\frac 12$ and $-\frac 12$), and
the spin contribution (resp. $+\frac 12$, $-\frac 12$, $+\frac 12$ and $-\frac 12$).

\begin{figure}
\includegraphics[scale=0.65]{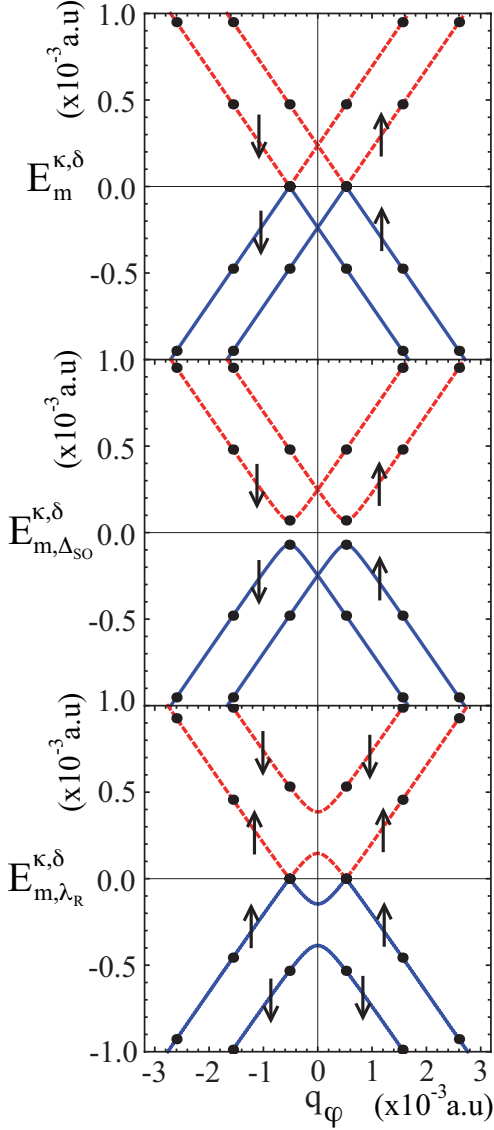}
\caption{Dispersion relations for metallic rings for the free (top panel) and both SO interactions (intrinsic middle and Rashba bottom, panels). The ISO has been drawn for $\DeltaSO=0.7 \times 10^{-4}{\rm a.u.}$, and opens a gap of size $2\DeltaSO$ with separate branches for each spin label. The Rashba interaction is depicted for $\lambdaSO=1.2\times 10^{-4}{\rm a.u.}$, the allowed values of $m$ are indicated by the full dots.  Note the spin asymmetry introduced by the Rashba coupling, that will have striking consequences for the charge and spin persistent currents.}
\label{Fig3}
\end{figure}

The eigenenergies, assuming these wave functions (with constant amplitudes $A$ an $B$), are 
\begin{equation}
\energy^{\kappa,\delta}_{m,\Delta}=\kappa\sqrt{\DeltaSO^2+\epsilon^2 (m-\delta/2)^2},
\end{equation}
where $\kappa=\pm 1$ is the particle-hole index and $\delta=\pm 1$ the SO index, and
$\epsilon=\frac{\hbar v_F}{a}$.  $\kappa\epsilon |(m-\delta/2)|$ corresponds to the 
electron energies in the absence of ISO. On the other hand when only the Rashba interaction is present
the energy is given by
\begin{widetext}
\begin{eqnarray}
\energy^{\kappa,\delta}_{m,\lambdaSO} =\frac{\kappa}{2}\sqrt{\epsilon^2(1+4m^2)+ 8\lambdaSO ^2 - 4\delta \sqrt{(m^2\epsilon^2+ \lambdaSO ^2)(\epsilon^2+4\lambdaSO^2),}}
\label{energia}
\end{eqnarray}
\end{widetext}
which has the correct zero SO coupling limit. These energies correspond to the angular wavectors satisfying the closed ring boundary conditions. The spectrum is shown in Fig.~\ref{Fig3}. We assume that the transverse mode is in the ground state using again as reference the transverse modes for the graphene  zigzag ribbons. The spinor wave functions for the ribbons depend on both longitudinal and transverse indices. Choosing the basis state in the $N\gg1$ limit permits writing an explicit expression for the wave functions and assess the coupling of the free transverse modes in the presence of the SO couplings. If the coupling is large compared to the transverse level separation, it must be contemplated in the analysis\cite{Egues}. 

Let us estimate, on the
basis of the previous considerations, the widths of the rings we are describing in the continuum approach: Independence of longitudinal and transverse modes for zig-zag boundary conditions is a good approximation when the width
of the nanoribbon is much larger that one primitive basis vector magnitude $a_0$, in length, as was shown
by the exact solutions in e.g. ref.\onlinecite{Enoki}. From this point of view the width of the ring has to be greater than $10 \times a_0$. 
The second issue is band mixing due to the SO coupling. This can be
estimated by calculating the energies of the transverse modes in nanoribbons
with zig-zag edges for the free case and then evaluating the magnitude of the matrix elements of the SO coupling
between these modes. 

The typical values used for intrinsic coupling are estimated in Ref.~\onlinecite{MinEtAl} using a microscopic tight-binding model with atomic spin orbit interaction. The Rasha interaction comes from the atomic spin orbit and Stark interactions and the intrinsic from the mixing between $\sigma$ and $\pi$ bands due to atomic spin orbit interaction. 
The coupling constants are given by the expressions,
\begin{eqnarray}
\DeltaSO&=&\frac{|s|\xi^2}{18(sp\sigma)^2}, \nonumber \\
\lambdaSO&=&\frac{e E z_0 \xi}{3(sp\sigma)}, \nonumber
\end{eqnarray}
where $|s|$ and $(sp\sigma)$ are hopping parameters in the tight-binding model, $s=-8.868$ eV and $(sp\sigma)=5.580$~eV, $\xi=6$~meV is the atomic SO strength of carbon, and $z_0\sim 3\times a_B$ ($a_B$ is the Bohr radius), is proportional to its atomic size. $\lambdaSO$ is proportional to the electric field, $E\approx 50~{\rm V}/300$ nm, perpendicular to the graphene sheet. This gives values for the SO parameters $\lambdaSO\approx 0.1$~K and $\DeltaSO\approx 0.01$~K.

The energies for different free transverse modes
for graphene and for zig-zag edges have been computed in ref.\onlinecite{Fertig}.
Their calculation is a function of the nearest neighbour hopping parameter $t=2.8$ eV. Taking their
results for the free case, the energy spacing between transverse modes for a ribbon
width of $8.66 \times a_0$ is $\sim 0.8$eV, for double this width ($17 \times a_0$) the energy
gap decreases to 0.42 eV. The matrix element of the SO couplings between the free states is bounded from above by their absolute magnitudes in graphene. The couplings for bare/suspended graphene, discussed above are $\Delta_{SO}\sim 0.569 \mu$eV and $\lambda_R=  6 \mu$eV, will not introduce any appreciable coupling between transverse modes. For the case of an enhanced SO due to hybridization to a substrate (Rashba SO) or edge functionalization (intrinsic SO) as we have discussed, the magnitude of the coupling reaches $100-200$ meV and brings it closer to the transverse mode gap, limiting the rings widths to below $20 \times a_0$. In conclusion, for the strongest SO coupling reported the rings are optimally described in the continuum for widths between $10-20 \times a_0$, while for smaller couplings the
with can be much larger within the radial ground state approximation.

Recently Shakouri et al\cite{Shakouri} have analysed rings with both Rashba and intrinsic Dresselhaus interactions (although not graphene), and consistently discussed the problem of the mixing of transverse (radial) states and the validity of the aforementioned considerations. They concluded that it is only when both interactions are present and of similar magnitude, that radial state mixing occurs so that at least two states have to be contemplated. Nevertheless when only one of these interaction is dominant, the single radial state approximation is valid. This will always be our
situation here.

Although the possible wave vectors take on discrete half integer values, they will trace a continuous change when a gauge field is applied. Close to the point of closest approach between the valence and conduction bands. For the ISO coupling these points are around $m=\pm1/2$ and the expansion takes the form
\begin{equation}
\energy^{\kappa,\delta}_{m,\Delta}=\kappa|\DeltaSO|+\frac{\kappa \epsilon^2 }{2|\DeltaSO|}(m\pm 1/2)^2+O\left((m\pm1/2)^4\right),
\end{equation}
while for the Rashba coupling the behavior is
\begin{eqnarray}
&&\energy^{\kappa,\delta}_{m,\lambdaSO}=\frac{\kappa |m\pm1/2|}{\sqrt{2}(\epsilon^2+4\lambdaSO^2)}+O\left((m\pm1/2)^2\right).\nonumber\\
\end{eqnarray}
The intrinsic spin-orbit term will open a gap in the vicinity of $(m=\pm1/2)$ which is simply $2\DeltaSO$ where the electrons exhibit an effective mass of $m^*_{\rm ISO}=\DeltaSO^2/v^2_F$ which is small, both because $v_F$ is large and $\DeltaSO$ is in the range of $\rm meV$ for graphene. For the Rashba coupling there is no gap at $m=\pm1/2$ but we will see a spin dependent gap opens continuously as the magnetic field is applied. Note also that this is a gap between spin-orbit up states. The gap between spin-orbit down states is given by $ \sqrt{\epsilon^2+4\lambdaSO^2}+2\lambdaSO$. One can define an effective mass of the spin down states as $m^*_{\downarrow}=\kappa\lambdaSO\hbar^2/[2\epsilon^2(2\lambdaSO+\sqrt{\epsilon^2+4\lambdaSO^2})]$. 

The limit in which the SO coupling goes to zero is singular, since both gaps close and the dispersion becomes linear as $\kappa \hbar v_F q_{\varphi}$. This limit highlights another feature of the Rashba spectrum; in the vicinity of the Dirac points $K$, and $K'$,  the electron behaves as a hole (has negative mass) in the conduction band and has negative charge (positive mass). From the expression above $m^*_{\uparrow}=-m^*_{\downarrow}$ at the Dirac point.


The split bands open a gap symmetrically between the $\delta$ states when $\DeltaSO=0$. If $\DeltaSO \neq 0$ the contributions for each gap are different~\cite{jaen,kue}. In this parametrization the blue and the red curves (dashed and continuous respectively) represent the levels in the quantization axis of the RSO interaction, i.e. in the SO basis\cite{yamamoto}. 

As we will see below, the velocity operator merits a non-trivial treatment in the context of graphene. For this reason we will derive the eigenfunctions for both SO couplings to compute the charge and spin persistent currents using the velocity operator, and compare it with the linear response relation. For the ISO only we have the wavefunctions
\begin{eqnarray}
\Psi^{\kappa,\delta}_{m,\Delta}(\varphi)&=&\frac{e^{i m\varphi}}{2 |E^{\kappa,\delta}_{m,\Delta}|}\begin{pmatrix}
\scriptstyle \delta_{\delta,+} [ \epsilon(m-\frac{1}{2})-i(\DeltaSO+E^{\kappa,\delta}_{m,\Delta})] e^{-i \varphi}\\
\scriptstyle \delta_{\delta,-}[\epsilon(m+\frac{1}{2})+i(\DeltaSO-E^{\kappa,\delta}_{m,\Delta})]
\phantom{e^{-i \varphi}}\\ 
\scriptstyle \delta_{\delta,+}[\epsilon(m-\frac{1}{2})-i(\DeltaSO-E^{\kappa,\delta}_{m,\Delta})]
\phantom{e^{-i \varphi}}\\
 \scriptstyle \delta_{\delta,+} [ \epsilon(m+\frac{1}{2})+i(\DeltaSO+E^{\kappa,\delta}_{m,\Delta}) ]e^{i\varphi}\phantom{^{-}}
 \end{pmatrix},\nonumber\\
\end{eqnarray}
labelled by $\kappa$ and $\delta$ as  $\Psi^{\kappa,\delta}_{m,\Delta}$. The polarization of this state is given by the expectation value of the operator $(\hbar/2)\nbOne_{\sigma}{\matbfs}$, 
\begin{equation}
\langle \mats_z\rangle=\frac{\hbar}{2}(\Psi^{\kappa,\delta}_{m,\Delta}(\varphi))^{\dagger}\nbOne_{\sigma}\mats_z\Psi^{\kappa,\delta}_{m,\Delta}(\varphi).
\end{equation}
and all the states are polarized perpendicular to the 
Corbino disk i.e. the $\bf z$ direction. This is also the direction of the effective magnetic field implied by the rewriting of the ISO term as $(\DeltaSO{\bm \sigma})\cdot {\matbfs}=(\DeltaSO\sigma_z)\mats_z$, a field that aligns the spins in opposite direction on different sublattices, in the $z$ direction. The result is zero global spin-magnetization while each sub lattice is spin-magnetized in opposite directions. This is in accordance with the fact that the intrinsic SO interaction operates as a local magnetic field in each sublattice with opposite sign, and thus not breaking of time reversal symmetry.

The pseudo spin polarizations are computed in an analogous fashion
\begin{eqnarray}
\langle {\bm \sigma}\rangle&=&\frac{\hbar}{2}(\Psi^{\kappa,\delta}_{m,\Delta}(\varphi))^{\dagger}{\bm \sigma}\nbOne_{s}\Psi^{\kappa,\delta}_{m,\Delta}(\varphi),\nonumber \\
&=&\frac{\kappa\hbar}{2}\frac{\delta \hat{\bm z}\Delta_{SO}+(m-\delta/2)\epsilon\hat{\bm\varphi}}{E_{m,\Delta}^{\kappa,\delta}},
\end{eqnarray}
where we note the ordering go the matrix direct product. One sees both orbital and spin-orbit contributions, 
so the pseudo spin does not simply follow the electron momentum.

The Rashba eigenfunctions are
\begin{eqnarray}
\Psi^{\kappa,\delta}_{m,\lambdaSO}(\varphi)&=&\frac{e^{im\varphi}}{\sqrt{\Lambda}}\begin{pmatrix}\frac{-2i E^{\kappa,\delta}_{m,\lambdaSO}(m\epsilon^2+2\lambdaSO^2+\delta\Gamma_m)}{(4m^2-1)\epsilon^2\lambdaSO}e^{-i\varphi}
\\ 
\frac{-2i E^{\kappa,\delta}_{m,\lambdaSO}}{\epsilon(2m+1)}
\\ 
\frac{m\epsilon^2-2\lambda^2+\delta\Gamma_m}{\epsilon \lambdaSO(2m+1)}
\\
e^{i\varphi}
\end{pmatrix},\nonumber\\
\end{eqnarray}
where $\Gamma_m=\sqrt{(m^2\epsilon^2+\lambdaSO^2)(\epsilon^2+4\lambdaSO^2)}$ and $\Lambda=4\Gamma_m\left(\Gamma_m-\delta(2\lambdaSO^2-m\epsilon^2)\right)/(2m+1)^2\epsilon^2\lambdaSO^2$.  The polarization of the Rashba eigenvectors is given by 
\begin{eqnarray}
\langle {\matbfs}\rangle&=&\frac{\hbar}{2}(\Psi^{\kappa,\delta}_{m,\lambdaSO}(\varphi))^{\dagger}\nbOne_{\sigma}{\matbfs}\Psi^{\kappa,\delta}_{m,\lambdaSO}(\varphi),\nonumber\\
&=&\delta\left(\frac{\hbar}{2}\right)\frac{m\epsilon(2\lambdaSO\hat{\bm \rho}+\epsilon\hat{\bf z})}{\Gamma_m},
\end{eqnarray}
where two contributions are evident, the polarization points outward in the radial direction and has a component due to the orbital rotation of the electrons.

Following previous expressions the Rashba pseudo-spin polarizations are
\begin{eqnarray}
\langle {\bm \sigma}\rangle&=&\frac{\hbar}{2}(\Psi^{\kappa,\delta}_{m,\lambdaSO}(\varphi))^{\dagger}{\bm \sigma}\nbOne_{s}\Psi^{\kappa,\delta}_{m,\lambdaSO}(\varphi),\nonumber \\
&=&\frac{\hbar}{2}\frac{\delta m\epsilon E_{m,\lambdaSO}^{\kappa,\delta}(\delta m\gamma+\Gamma_m)\hat{\bm\varphi}}{(m-1/2)(\delta m\epsilon^2\Gamma_m+m^2\epsilon^2\gamma+\lambdaSO^2(\gamma-2\delta\Gamma_m))},\nonumber\\
\end{eqnarray}
where $\gamma=\epsilon^2+4\lambdaSO^2$.

\section{Charge persistent currents}
Persistent equilibrium currents are a direct probe of energy spectrum of the system in the vicinity of the Fermi energy. Although such currents are typically small and are detected by the magnetic moment they produce\cite{VonOppen}, recent experiments, where many rings form dense arrays on a cantilever, boost the magnetic signal allowing both measurement of the current signal and the use of the set up as a sensitive magnetometer. The Corbino disk geometry can be easily built with high precision by using new techniques\cite{flakecuttingpaper} manipulating nano-particles as cutters and hydrogenating the open bonds.
\begin{figure*}
\centering
\includegraphics[width=14.6cm]{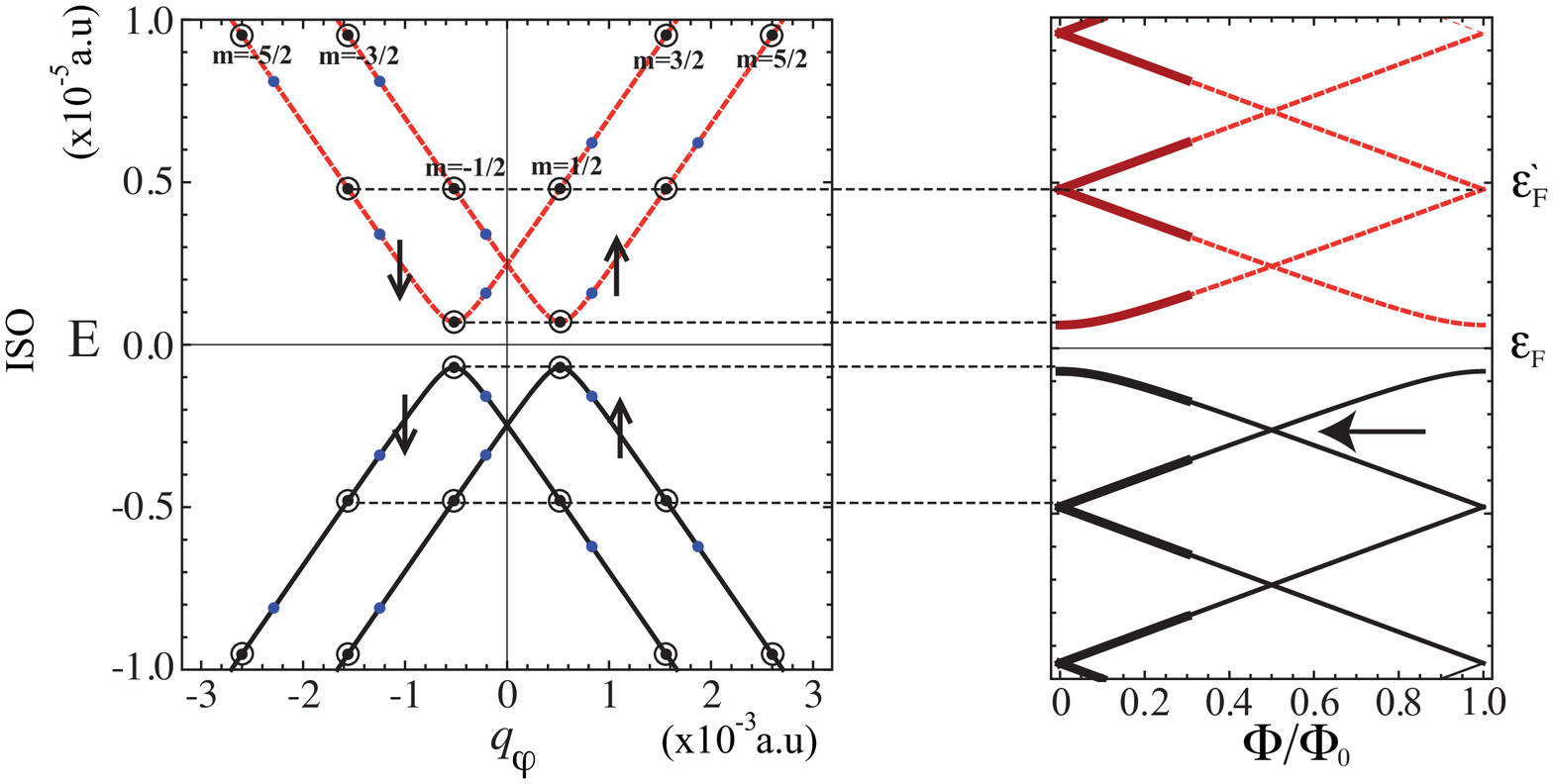}  
\includegraphics[width=14.4cm]{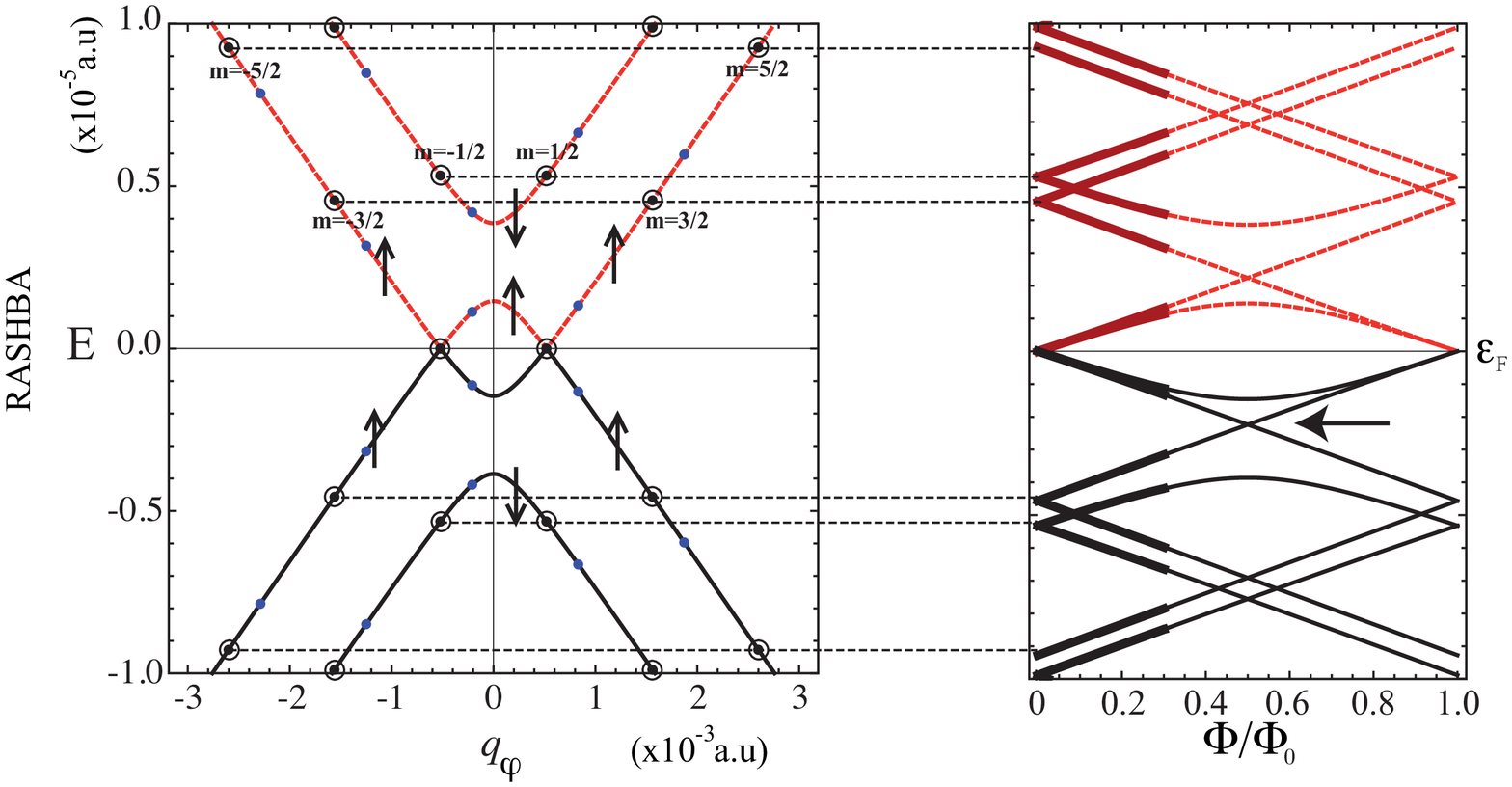}  
\caption{The energy dispersion for the ISO (top panel with $\DeltaSO=7\times 10^{-5}$ a.u. see ref.(\onlinecite{MinEtAl})) and Rashba coupling (bottom panel with $\lambdaSO=1.2\times 10^{-4}$a.u.) as a function of wave-vector $q_{\varphi}$ (for continuum range of values; solid and dashed lines). The circled dots represent the allowed values of the energies on the ring at zero magnetic flux. The uncirled dots (blue online) represent the shift of the allowed energies due to a finite flux. On the right panel, the energy bands are plotted against $\Phi/\Phi_0$. On this panel the trajectory of the allowed values of the energy is followed as a function of the field. The solid lines represent the valence bands and the dashed the conduction bands. The Fermi energy is assumed to be zero, except for the Rashba where a finite value for the Fermi energy is also illustrated. The bold arrow on the right panels indicate level crossing discussed in the text.} 
\label{f:figureEMF-AC1}
\end{figure*}

The spectrum of the system is modified by a field flux perpendicular to the Corbino disk as follows
\begin{widetext}
\begin{eqnarray}
\energy^{\kappa,s}_{m,\Delta}(\Phi)&=&\kappa  \sqrt{\DeltaSO^2+\epsilon^2 (m-\delta/2+\Phi/\Phi_0 )^2},\\
\energy^{\kappa,\delta}_{m,\lambdaSO}(\Phi) &=&\frac{\kappa}{2} \sqrt{8\lambdaSO ^2+\epsilon^2\left( 4(m+\Phi/\Phi_0 )^2+1\right)- 4\delta \sqrt{\left(4\lambdaSO ^2+\epsilon^2\right)
   \left(\lambdaSO ^2+\epsilon^2 (m+\Phi/\Phi_0)^2\right)}},
\label{energiaFlux}
\end{eqnarray}
\end{widetext}
where the  Zeeman coupling has been neglected at small enough fields. The addition of a magnetic field, in the form of a $U(1)$ minimal coupling with flux $\Phi$ threading the ring, breaks time reversal symmetry allowing for persistent charge currents\cite{ImryButtiker}. In the case of a ring of 
constant radius threaded by a perpendicular magnetic flux, the angular component of the gauge vector $A_\varphi=\Phi/2\pi a$ may be eliminated via a gauge transformation $A'_\varphi=A_\varphi+a^{-1}\partial_\varphi\chi=0$, $\Psi'(\varphi)=\Psi(\varphi)e^{ie\chi/\hbar}$ at the expense of modifying the BCs on the ring to
\be\Psi'(\varphi+2\pi)=e^{-i\theta_0}e^{-2i\pi\Phi/\Phi_0}\Psi'(\varphi),\label{eqModifiedBC}\ee
where $\Phi_0$ is the normal quantum of magnetic flux $(h/e)$. As mentioned before, the twist in the BCs and the field accomplish the same effect, so one can use them interchangeably
while satisfying the relation
\be
\energy^{\kappa,\delta}_{m,\Phi}(\theta_0)=\energy^{\kappa,\delta}_{m,0}(\theta_0+2\pi\Phi/\Phi_0),
\ee
hence $m\rightarrow m-\theta_0/2\pi+\Phi/\Phi_0$, as discussed in Eq.\ref{EqEnergiesTwistedGraphene}.
The energy dispersion for the graphene ring is illustrated in Fig.~\ref{f:figureEMF-AC1} (left panel), where the different colors (online) (see caption) refer to the conduction band ($\kappa = +1$, dashed line) and valence band ($\kappa = -1$, full line). As expected, the energy levels display a periodic variation with the magnetic flux (right panel in the figure). 

The charge persistent current in the ground state can be derived using the linear response definition $J_Q=-\sum'_{m,\kappa,\delta} \frac{\partial \energy}{\partial \Phi}$, where the primed sum refers to all occupied states only.  Since the current is periodic in $\Phi/\Phi_0$ with a period of 1, we can restrict the discussion to the window $0\le\Phi <\Phi_0$ where the occupied states are in the valence band $\kappa = -1$, since the Fermi level is chosen at the zero of energy.  We will first discuss the simple ISO coupling. The analytical expression is given by
\begin{equation}
J^{\kappa}_{Q,\Delta}=-\frac{\epsilon^2\kappa}{\Phi_0}\sum'_{m,\delta}\frac{(m-\delta/2+\Phi/\Phi_0)}{E^{\kappa,\delta}_{m,\DeltaSO}(\Phi)}.
\label{chargecurrentISO}
\end{equation}
In Fig.\ref{f:figureEMF-AC1}, on the left panel, the spin-orbit branches of the spectrum labeled with their spin quantum number have been depicted. The encircled dots are the allowed energy values, due to quantization on the ring, at zero magnetic field. When the field is turned on, these dots are displaced (no longer encircled) on the energy curve.
 
\begin{figure}
\centering
\includegraphics[scale=0.58]{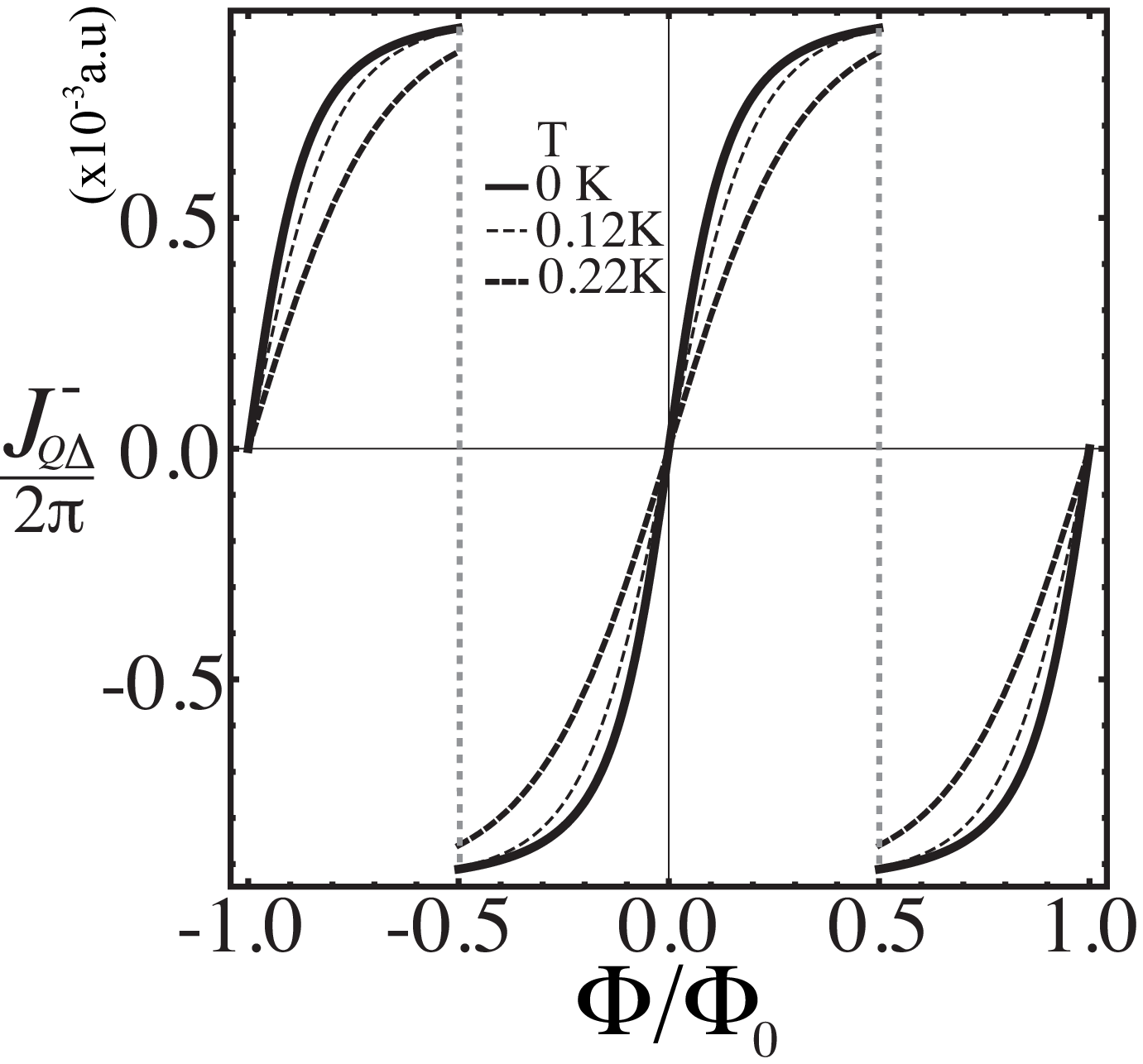}
\includegraphics[scale=0.58]{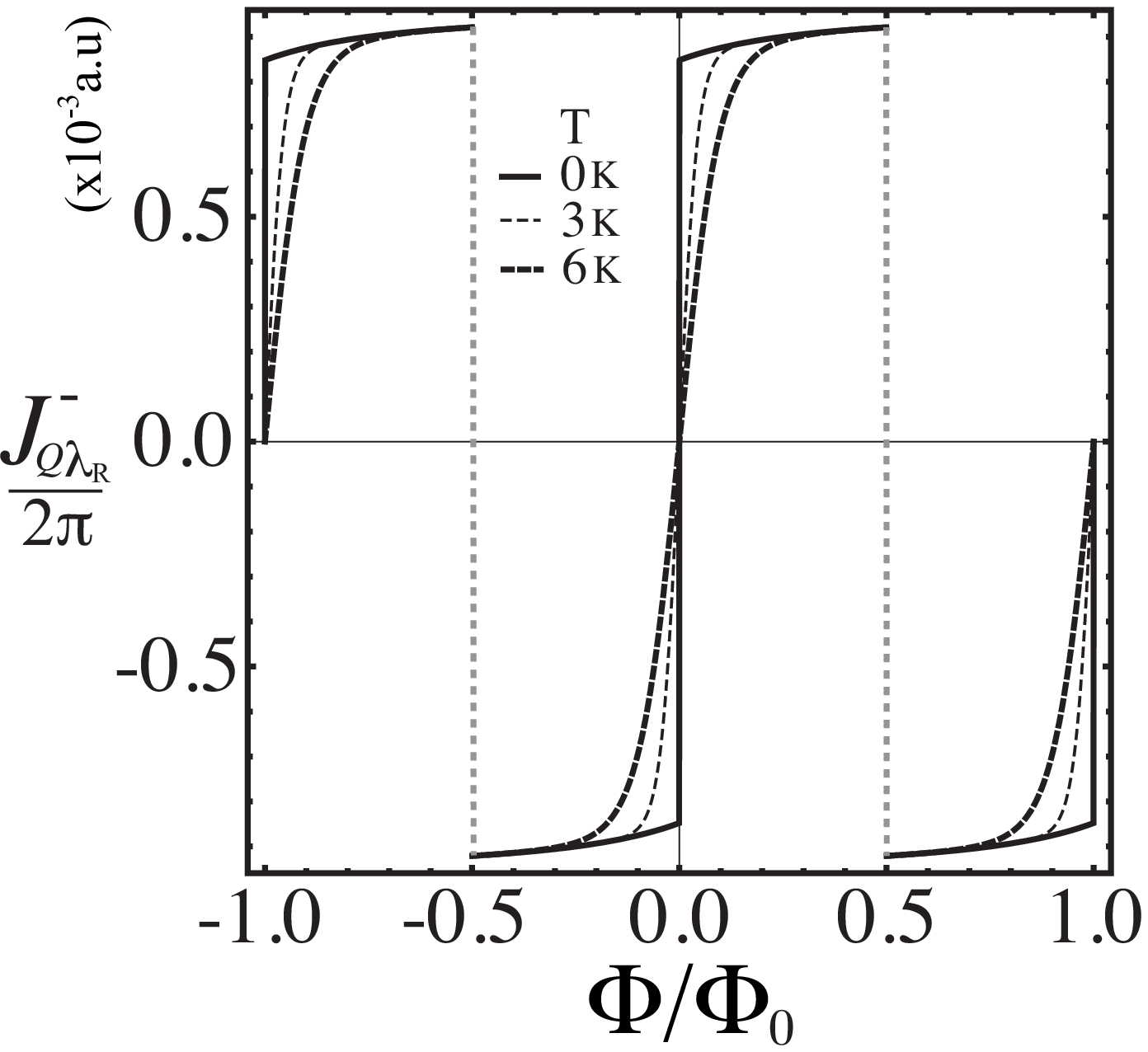}
\caption{ The equilibrium charge currents for both ISO and Rashba interactions. The ring is considered to have a radius of $a =20~$nm, with the same SO couplings as previous figures. The variations of the current are given by the slopes of the Figure \ref{f:figureEMF-AC1}. At each flux region, the states up to $\varepsilon_F$ are taken into account.}
\label{f:figureSOC}
\end{figure}

On the right panel we depict the trajectory of these dots as the magnetic field is increased for both the filled (full lines in figure) and unfilled (dashed lines) states. The negative derivative of the curves on the right panel added over the occupied states (both spin quantum numbers) is the net charge persistent current. For the range of energies shown, the only net contribution is from the levels closest and below the Fermi level. The lower levels have currents that tend to compensate in pairs. Following the curve on the right, below the Fermi energy and from zero field, the current first increases linearly and then bends over to reach a maximum value before two levels cross (crossing indicated by arrow on the right panel of Fig.\ref{f:figureEMF-AC1}). At that point, one follows the level closest to the Fermi energy (from below), the current changes sign and increases crossing the zero current level, whereupon the whole process repeats periodically. Such behavior is shown in Fig.\ref{f:figureSOC} top panel. Changing the Fermi level can change the scenario qualitatively. For example adjusting the Fermi level to $\varepsilon'_F$ (see Fig.\ref{f:figureEMF-AC1}), the currents would follow a square wave form, alternating between constant current blocks of opposite signs. 

For the Rashba coupling, represented in the bottom panels in Fig.\ref{f:figureEMF-AC1}, the current is derived in a similar way, but now there is a striking asymmetry between spin branches. The analytical form for the charge current is
\begin{eqnarray}
&&J^{\kappa}_{Q,\lambdaSO}=\nonumber \\
&&-\frac{\epsilon^2\kappa}{\Phi_0}\sum'_{m,\delta}\frac{\left (2-\frac{\delta(\epsilon^2+4\lambdaSO^2)}{\sqrt{(\epsilon^2+4\lambdaSO^2)(\lambdaSO^2+\epsilon^2(m+\Phi/\Phi_0)^2)}}\right )(m+\Phi/\Phi_0)}{E^{\kappa,\delta}_{m,\lambdaSO}(\Phi)}.\nonumber\\
\end{eqnarray}
The spin branch closest to the Fermi energy is non monotonous, making for two different contributions
to the charge current for the up spin contribution. Note also that we have taken into account the current coming
from the spin down branch which does not have the same effective mass as the corresponding branch of
the opposite spin. The results are depicted in Fig.\ref{f:figureSOC} bottom panel. The structure of the spectrum
being asymmetric between spin branches makes for the possibility of net spin currents as we will see below.
The charge persistent current can be manipulated with $\lambdaSO$ since the Rashba parameter can be tuned by a field perpendicular to the plane of the ring. In contrast, the intrinsic SO cannot
be easily tuned by applying external fields. Nevertheless, it has been established experimentally\cite{Balakrishnan} that light covering of graphene with covalently bonded hydrogen atoms modifies the carbon hybridization and can enhance the intrinsic spin-orbit strength by three orders of magnitude\cite{Balakrishnan}. Regulating this covering may then be a tool to manipulate charge currents.

One can contemplate the effect of temperature on the robustness of persistent charge currents by considering the occupation of the energy levels. The Fermi function has then to be factored into the computation of the currents
\begin{equation}
J^{\kappa}_{Q,\lambdaSO}(T)=-\sum_{m,\delta}\frac{\partial \energy^{\kappa,\delta}_{m,\lambdaSO}(\Phi)}{\partial\Phi}f(\energy^{\kappa,\delta}_{m,\lambdaSO},\varepsilon_F,T),
\end{equation}
where 
$f(E
,\varepsilon_F,T)=
\left (1+\exp{(E
-\varepsilon_F)/k_B T}\right )^{-1}$ is the Fermi occupation function for the case of the Rashba coupling. There is no need now to restrict the energy levels contemplated since the filling is determined by the Fermi distribution.

Figure \ref{f:figureSOC},  shows the effect of a temperature energy scale of the order of the SO strength for both intrinsic and Rashba couplings. The deep levels will be fully occupied while the shallow levels (close to the
Fermi energy) will have a temperature dependent occupancy. Occupation depletion affects mostly the current contributions from levels within $k_B T$ of the Fermi level. This typically happens in the vicinity of the integer values of the normalized flux $\Phi/\Phi_0$, but at half integer fluxes  the contributing levels dig into the Fermi sea where carrier depletion is less pronounced and current discontinuities tend to be protected from temperature effects. From Fig.\ref{f:figureEMF-AC1} one can estimate the depth in energy of the crossing to be $\sim 3\times 10^{-4}$ a.u which amounts to a temperature equivalent of $\sim$1 K before degradation of spin currents is observed at half integer fluxes. This is an important feature of the linear dispersions in graphene, and in enhanced SO coupling scenarios could be of applicability for magnetometer devices at relatively higher temperatures.

\section{Equilibrium spin currents}
We now contemplate spin equilibrium currents. In the absence of a direct linear response definition one
can obtain them from the charge currents by distinguishing the velocities of different
spin branches. We define a spin equilibrium current as
\be
J_{\rm S}=J_Q(\delta=-1)-J_Q(\delta=1), 
\ee
where one weighs the asymmetry in velocities of the different occupied spin branches. 
As we mentioned in the previous section there is no spin asymmetry both for the free case and for the ISO, so no spin current can result in this case, i.e. both spin branches contribute charge current with the same amplitude so
they cancel in the above expression. With the Rashba coupling, the inversion symmetry is broken
inside the plane and the spin branches are asymmetrical for a range of $q_{\varphi}$ values. 
\begin{figure}
\includegraphics[scale=0.55]{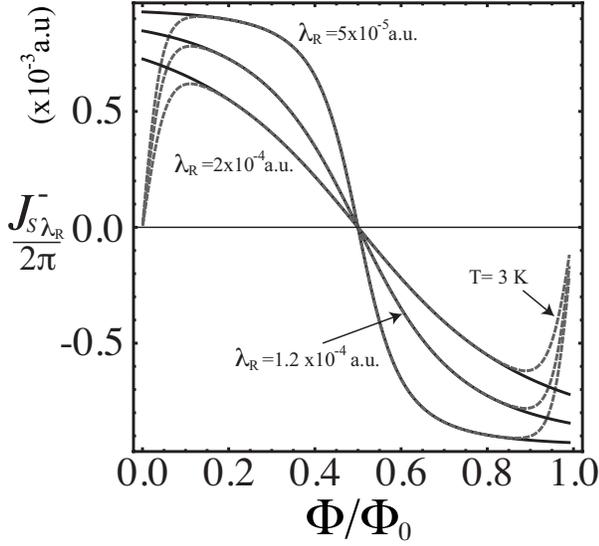}
\caption{Spin current for the Rashba SO coupling indicated by the legends as a function of the magnetic field, as derived form the charge currents distinguished by spin components. The large spin currents at small fields are due to dominant charge currents for a single spin orientation. Temperature affect currents at the integer values of flux, while toward haslf integer values currents are protected.}
\label{figSpinCurrent}
\end{figure}

The peculiar separation of the spin branches makes for velocity differences of the two spin projections and a
spin current ensues as shown in Fig.\ref{figSpinCurrent}. The figure shows a large spin current for small fluxes that can be traced back to the large charge currents coming from a single spin branch in Figure \ref{f:figureEMF-AC1}. Toward half integer flux quantum's the opposite spin charge current increases until it cancels out the spin current completely. Beyond half integer flux the spin current is reversed in sign and at zero temperature there is a discontinuity approaching integer fluxes. As discussed for charge currents, the spin currents are also most susceptible to thermal depletion of carriers at integer fluxes, while toward half integer fluxes these are protected. 

A striking feature, that survives temperature effects, is that the spin currents increase as one lowers $\lambdaSO$. The Rashba coupling breaks inversion symmetry in the plane even for small $\lambdaSO$. The symmetry breaking determines the spin labeling of the energy branches that take part in the spin current. It is only for $\lambdaSO=0$ that the free Hamiltonian symmetry is re-established and the spin currents are destroyed. A combination of the described symmetry effect and the thermal shielding from deep levels make for these effects observable experimentally. 

\section{Velocity operators for graphene}
As discussed in section III, there are two ways to compute the effect of the magnetic field: either putting the description
in the Hamiltonian as a gauge vector or performing a gauge transformation and passing all field information to the wave function.
For SU(2) gauge theory applied to the present case, this process cannot be done directly because of the lack of
gauge symmetry\cite{BercheMedinaLopez}. We have solved the problem fully for the ``gauge fields" in the
Hamiltonian and determined the eingenfunctions. Such eigenfunctions contain the full information of the
state, and the velocities as a function of the magnetic field can be derived by using the canonical equations
$v_\varphi=a\dot\varphi=\frac {a}{i\hbar}[\varphi,H]$ where the commutator takes the value
$[\varphi,H]=i\hbar v_F a^{-1}\bsigma_\varphi\nbOne_s$ and compute
\begin{eqnarray}
J_Q=\frac{ev_F}{a}\sum'_{m,\delta} \left(\Psi_{m}^{\kappa,\delta}(\varphi,\Phi)\right)^{\dagger} \bsigma_\varphi\nbOne_s\Psi_{m}^{\kappa,\delta}(\varphi,\Phi).
\label{velocitywrong}
\end{eqnarray}
Taking the ISO wave functions and substituting $m\rightarrow m+\Phi/\Phi_0$ we determine the appropriate $\Psi_{m}^{\kappa,\delta}(\varphi,\Phi)$.
We could also, leave the wave function untouched and include a U(1) gauge vector in the momentum operator. Let us explicitly
write out an expectation value
\begin{widetext}
\begin{eqnarray}
&&J^{+,+}_{Q,\Delta}=-ev_F\left(\Psi_{m,\Delta}^{+,+}(\varphi,\Phi) \right)^{\dagger}\sigma_{\varphi}\nbOne_s\Psi_{m,\Delta}^{+,+}(\varphi,\Phi)=\nonumber\\
&&\frac{1}{4 |E^{+,+}_{m,\Delta}|^2}\begin{pmatrix}\left [ \epsilon(m-\frac{1}{2}+\frac{\Phi}{\Phi_0})+i(\DeltaSO+E^{+,+}_{m,\Delta})\right ] e^{i \varphi}\\0\\ \epsilon(m-\frac{1}{2}+\frac{\Phi}{\Phi_0})+i(\DeltaSO-E^{+,+}_{m,\Delta})\\0\end{pmatrix}^T \begin{pmatrix} 0&-ie^{-i\varphi}\\ie^{i\varphi}&0\end{pmatrix}\otimes\begin{pmatrix}1&0\\0&1\end{pmatrix}\begin{pmatrix}\left [ \epsilon(m-\frac{1}{2}+\frac{\Phi}{\Phi_0})-i(\DeltaSO+E^{+,+}_{m,\Delta})\right ] e^{-i\varphi}\\0\\ \epsilon(m-\frac{1}{2}+\frac{\Phi}{\Phi_0})-i(\DeltaSO-E^{+,+}_{m,\Delta})\\0
\end{pmatrix}\nonumber\\
&&=-\frac{\epsilon^2}{\Phi_0}\frac{\left (m-1/2+\Phi/\Phi_0\right )}{E^{+,+}_{m,\Delta}},
\end{eqnarray}
\end{widetext}
which coincides with the expression of Eq.\ref{chargecurrentISO}.
With either of the two procedures one retrieves the same charge current of Eq.\ref{chargecurrentISO}. This is a simple but interesting connection
between linear response relations used to compute the current and a canonical exact calculation in principle. Note also that this expectation value corresponds to the procedure that eliminates Zitterbewegung from the Dirac definition of the velocity operator $\langle c{\bm\alpha}\rangle$ where ${\bm \alpha}=\sigma_\varphi\nbOne_s$ and $c=v_F$.  One can also obtain the linear response result using the group velocity operator applied to the free wave functions\cite{Baym}, where the group velocity operator is then
\begin{equation}
{\hat J}^{\kappa,\delta}_{Q,\Delta}=\frac{v^2_F {\bf {\hat p}}}{\kappa\sqrt{\DeltaSO^2+\epsilon^2 (m-\delta/2)^2}},
\end{equation}
where ${\bf {\hat p}}=(-i\hbar/a)\partial_\varphi$. The first procedure above does not work for the Rashba coupling, that is, sandwiching the ordinary velocity operator in between the Rashba wave functions does not yield the linear response result. The second, group velocity approach depends on finding an appropriate Foldy-Wouthuysen transformation we believe is not currently known in the literature. These issues remain topic for future work.

\section{Summary and Conclusions}
We have discussed equilibrium currents in a Corbino graphene ring, taking into 
account Rashba and intrinsic spin-orbit couplings separately. The ring is threaded 
by a magnetic flux and an electric field perpendicular to the graphene surface in 
order to tune the Rashba coupling. A detailed discussion was given, for setting up the correct 
Hamiltonian in polar coordinates and for the spinor wave functions closure
conditions. Twisted boundary conditions are discussed as a gauge freedom
useful in our treatment where the magnetic flux can be translated from the 
Hamiltonian to the wave function. Four quantum numbers are necessary to 
describe the energy eigenvalues, the valley index $\tau$ the particle hole index $\kappa$, 
the spin-orbit quantum number $\delta$,  labeling the spin quantization axis and the
angular momentum quantum number $m$.

The width of the rings, describable in terms of a continuum description
including generic zig-zag boundaries, assuming only the ground radial
state of the ring, were discussed. Our approach is valid for Corbino ring widths between
at least 10-20 times the magnitude of the primitive lattice vectors.
The upper limit is determined by the radial state gap for the free case, the
possible width of the ring increasing as the SO coupling is reduced.

The charge equilibrium currents are directly calculated from the spectrum, using linear
response relations, for small magnetic fluxes (so the Zeeman coupling can be neglected) and as a function of the
spin-orbit couplings.  We were able to derive an explicit simple form for the four spinor
in the case of zero Rashba interaction. The charge currents are induced by the magnetic flux,
as expected. While spin-orbit interactions do not induce charge currents by themselves 
(they preserve time reversal symmetry) we showed that at a non-zero fixed flux, away 
from $\pm h/2e$, they can modify the charge current. This is done through the 
Rashba coupling that can be varied by gate voltages in the Corbino geometry.

Temperature effects have been addressed to determine whether persistent currents computed here are robust
at experimentally accessible conditions. The equilibrium current turn out to be more temperature
sensitive in the vicinity of integer flux, while for half-integer flux (where they are the largest) 
the currents are protected because they arise from contributions of levels submerged in the Fermi sea.   For the SO
strengths considered, equilibrium currents would be strong even at temperatures close to 1K.

Finally, we derived equilibrium spin currents on the Corbino disk, by combining charge current contributions
from opposite spin-orbit labels. Spin currents only exist for Rashba type SO coupling (they cancel
exactly of ISO interactions) and they exhibit the same temperature dependence 
as the charge currents, but in contrast, they are the more robust when their magnitude is smaller. 
A brief discussion was made regarding alternative definitions of equilibrium currents that are only successful 
for ISO type interactions. Analogous formulations for Rashba interactions are left for future work.

\acknowledgments
The authors acknowledge funding from the project PICS-CNRS 2013-2015. N.B. acknowledges ``Coll\`ege Doctoral Franco-Allemand 02-07'' for financial support.

\end{document}